\documentstyle[preprint,aps,eqsecnum]{revtex}

\def\beq#1{\begin{equation} \label{#1}}
\def\eeq{\end{equation}}
\newcommand{\bea}{\begin{eqnarray}}
\newcommand{\eea}{\end{eqnarray}}
\def\bra#1{\left\langle #1\right\vert}
\def\ket#1{\left\vert #1\right\rangle}
\def\epsp{\epsilon^{\prime}}                    
\def\NPB{{ Nucl. Phys.} B}
\def\PLB{{ Phys. Lett.} B}
\def\PRL{ Phys. Rev. Lett.}
\def\PRD{{ Phys. Rev.} D}
\def\AJP{{\em Am. J. Phys.}}
\begin{document}
{
\tighten

\title {What is coherent in neutrino oscillations }
\author{Harry J. Lipkin\,\thanks{Supported
in part by grant from US-Israel Bi-National Science Foundation
and by the U.S. Department
of Energy, Division of High Energy Physics, Contract W-31-109-ENG-38.}}
\address{ \vbox{\vskip 0.truecm}
  Department of Particle Physics
  Weizmann Institute of Science, Rehovot 76100, Israel \\
\vbox{\vskip 0.truecm}
School of Physics and Astronomy,
Raymond and Beverly Sackler Faculty of Exact Sciences,
Tel Aviv University, Tel Aviv, Israel  \\
\vbox{\vskip 0.truecm}
High Energy Physics Division, Argonne National Laboratory,
Argonne, IL 60439-4815, USA\\
~\\harry.lipkin@weizmann.ac.il
\\~\\
}

\maketitle

\begin{abstract} 

Simple rigorous quantum mechanics with no hand waving nor loopholes clarifies
the confusion between three contradictory descriptions of the coherence
between  different neutrino mass eigenstates that can give rise to 
oscillations: (1) The standard textbook description of oscillations in time
produced by coherence between states with different masses and different
energies. (2) Stodolsky's proof that interference between states having
different energies cannot be observed in realistic experiments.  (3) The
description of a pion decay at rest into an observed muon and unobserved
neutrino as a ``missing mass" experiment where coherence between  different
neutrino mass eigenstates is not observable.

The known position in space of all realistic  detectors is rigorously shown to 
provide the quantum-mechanical ignorance of the neutrino momentum needed to
produce coherence between amplitudes from neutrino states with the same energy
and different masses.. Conditions  are precisely formulated for the loss of
coherence when mass eigenstate wave packets moving with different velocities
separate. The example of Bragg scattering shows how quantum-mechanically
imposed ignorance produces coherence.

\end{abstract}

} 


\def\beq#1{\begin{equation} \label{#1}}
\def\eeq{\end{equation}}
\def\bra#1{\left\langle #1\right\vert}
\def\ket#1{\left\vert #1\right\rangle}
\def\epsp{\epsilon^{\prime}}
\def\NPB{{ Nucl. Phys.} B}
\def\PLB{{ Phys. Lett.} B}
\def\PRL{ Phys. Rev. Lett.}
\def\PRD{{ Phys. Rev.} D}
\section{ How Neutrinos with different masses can be Coherent}

\subsection{Introduction} 

The standard textbook description shows that a coherent linear combination of
neutrino eigenstates with the same momentum and different masses  have
different energies and oscillate in time. But  such time oscillations and
coherence between states having different energies are not observed in most
realistic experiments\cite{Pnonexp,Leo}. Furthermore  coherence or
interference  between different neutrino mass eigenstates cannot be observed in
a ``missing mass" experiment where the mass of an unobserved  neutrino is
unqiuely determined by other measurements and momentum and energy conservation.

The resolution of these contradictions is just simple quantum mechanics. In any
experiment which can detect neutrino oscillations, the position of the detector
must be known with an error much smaller than the wave length of the
oscillation to be observed. The quantum mechanical uncertainty principle
therefore forces coherence between neutrino mass eigenstates having the same
energy and different momenta. Time behavior, time measurements and stationarity
in energy\cite{Leo} are irrelevant for this conclusion. The location in space
already says it all.

This simple physical argument is now spelled out rigorously with simple quantum
mechanics and no hand waving. In all realistic experiments the product of the
quantum fluctuations in the position of the detector and the momentum range
over which coherence is established is a very small parameter. Expanding the
exact transition matrix element for the neutrino detection in powers of this
small parameter and taking the leading term gives the desired  result.

\subsection{No coherence in a missing mass experiment}

When a pion decays at rest $\pi \rightarrow \mu \nu$ the energies $E_\mu,E_\nu$
and momenta  $\vec p_\mu, \vec p_\nu$ of the neutrino and muon can all
be known. This is just a ``Missing Mass" experiment. The neutrino mass $M_\nu$ is
uniquely determined by $M_{\nu}^2 = (M_{\pi} - E_{\mu})^2 - p_{\mu}^2$. So how
can there be coherence and interference between states of different mass? We
are guided to the resolution  of this paradox by experience in condensed matter
physics discussing which amplitudes are coherent in quantum
mechanics\cite{Frank,Weber,ADY,Kayser}.

The original Lederman-Schwartz-Steinberger experiment found that the neutrinos
emitted in a $\pi-\mu$ decay produced only muons and no electrons. Experiments
now show that at least two neutrino mass eigenstates are emitted in $\pi-\mu$
decay and that at least one of them can produce an  electron in a neutrino
detector. The experimentally observed absence of electrons can be explained
only if the electron amplitudes received at the detector from different
neutrino  mass eigenstates are coherent and exactly cancel. This implies  that
sufficient information was not available to determine the  neutrino mass from
energy and momentum conservation. A missing mass experiment was not  performed.

\subsection{Why quantum-mechanically imposed ignorance is needed}

Destruction of information by simple ignorance or stupidity cannot provide
coherence. The experimental setup must forbid via the  quantum-mechanical
uncertainty principle the knowledge of the information necessary to determine
the neutrino mass.  This paper analyzes the basic physics and presents a
rigorous quantitative analysis of the hand-waving uncertainty principle
argument. The knowledge of the position of any realistic neutrino detector is
shown to be  sufficient to provide the uncertainty in momentum needed to create
coherence between the amplitudes carried to the detector by components in the
neutrino wave function with the same energy, different masses and different
momenta.  

The initial state of the detector before the interaction with the neutrino is
described by a many-body wave function that exists only in a finite region of
configuration space. The probability is zero for finding any detector nucleon
anywhere in space outside of this volume. This exact property of the exact
initial state is rigorously shown below to prevent the detector  from
recognizing the difference between two incident neutrinos with the same energy
and slightly different momenta. It ensures the  quantum-mechanical ignorance
needed to produce coherence.  This physics can  be handwaved and called the
uncertainty principle. But it can also be proved
rigorously\cite{pwhichfin}.

\section{The Basic Physics of Neutrino Detection}

\subsection {The neutrino wave packet}

The neutrino wave packet traveling between source and detector 
vanishes outside some finite interval in space at any given time. At any point
on its path it also vanishes outside some finite time interval. 
The packet therefore contains components with different momenta and different
energies which are all coherent with well  defined phases to cancel out at all
points in space and time where the probability of finding the neutrino
vanishes.   

However, not all the different kinds of coherence present in the wave packet
are observable with a conventional detector. The detector is sensitive in very
different ways to the different components in the wave packet.
\cite{Leo,pwhichfin,grimus}.

\subsection {The role of the neutrino detector} 

Neutrino absorption is a weak interaction described completely by
the transition matrix of the weak interaction operator between the exact
initial and final states of the lepton and detector, where the exact states
include all strong interactions. This matrix element can be expanded in
powers of a small parameter, the product of the displacement of the
detector nucleon from the center of the detector and a momentum interval
which includes all momenta of incident neutrinos having the same energy.

We shall now show that
the leading term in this expansion gives the lepton flavor output for each
energy component in the initial neutrino wave function as the coherent sum of
the contributions from states with the same energy and different momenta. This
is exact subject only to corrections of higher order in the small parameter
which are negligible as long as the size of the detector is negligibly small in
comparison with any neutrino oscillation wave length.

Consider the transition matrix element between an initial state $\ket{i(E)}$
with energy $E$ of   the entire neutrino - detector system and a final state
$\ket{f(E)}$ of the system of a charged muon and the detector with the same
energy E, where a neutrino $\nu_k$ with energy, mass and momentum $E_\nu$,
$m_k$ and  $\vec P_o + \vec {\delta P_k}$ is detected via the transition 

\beq{weak1}
\nu_k + p \rightarrow \mu^+ + n  
 \eeq
occurring on a proton in the detector.  We express the neutrino momentum
as the sum of the mean momentum $\vec P_o$ of all the neutrinos  
with energy
$E_\nu$ and the difference $\vec {\delta P_k}$ between the momentum of each 
mass eigenstate and the mean momentum,

The transition matrix element depends upon the individual mass eigenstates $k$
only in the momentum difference $\vec {\delta P_k}$ and a factor $c_k$ for each
mass eigenstate which is a function of neutrino mixing angles describing the
transition amplitude for this mass eigenstate to produce a muon when it reaches
the detector. The transition matrix element can thus be written in a 
factorized form with one factor $T_o$ independent of the mass $m_k$ of the
neutrino  and a factor depending on $m_k$. 

\beq{weak2} \bra{f(E)}T \ket{i(E)} =  \sum_k
\bra{f(E)}T_o \cdot c_k e^{i\vec {\delta P_k}  \cdot \vec X} \ket{i(E)}  
\eeq   
where $\vec X$ denotes the co-ordinate of the nucleon that absorbs the
neutrino. Then if the product $\vec {\delta P_k}  \cdot \vec X$ of the momentum
spread in the neutrino wave packet and the fluctuations in the position of the
detector  nucleon is small,  the exponential can be expanded and approximated
by the leading term

\beq{weak3}
\bra{f(E)}T \ket{i(E)} = 
\sum_k \bra{f(E)}T_o \cdot c_k e^{i\vec {\delta P_k}  \cdot \vec X} \ket{i(E)} 
\approx
\sum_k \bra{f(E)}T_o \cdot c_k \ket{i(E)}  
 \eeq

The transition matrix element for the probability that a muon is observed at
the detector is thus proportional to the coherent sum of the amplitudes $c_k$
for neutrino components with the same energy and different masses and 
momenta to produce a muon at the  detector. A similar result is obtained for
the probability of observing each other flavor.
The final result is obtained by summing the contributions over all the energies
in the incident neutrino wave packet. But as long as the flavor output for each
energy is essentially unchanged over the energy region in the wave packet, the
flavor ouput is already determined for each energy, and is independent of any
coherence or incoherence between components with different energies. 

For  the case of two neutrinos with energy $E$ and mass eigenstates
$m_1$ and $m_2$ the relative phase of the two neutrino waves at a distance $x$
is:
\beq{WW3a}
\phi^E_m(x)= (p_1 - p_2)\cdot x =
{{(p_1^2 - p_2^2)}\over{(p_1 + p_2)}}\cdot x  =
{{\Delta m^2}\over{2p}}\cdot x 
\eeq
where $\Delta m^2 \equiv m_2^2-m_1^2$, and we have assumed the free space
relation between the masses, $m_i $ energy $E $ and momenta: 

The flavor output of the detector is thus seen to be determined by the
interference between components in the neutrino wave paclet with the same
energy and different masses and momenta.   All the relevant physics is
in the initial state of the nucleon in the detector that detects the
neutrino and emits a charged lepton, together with the relative phases of
the components of the incident neutrino wave packet with the same energy.

This result (\ref{weak3}-\ref{WW3a}) is completely independent of the neutrino
source and in particular completely independent of whether the source satisfies
Stodolsky's stationarity condition\cite{Leo}. No subsequent time measurements
or additional final state interactions that mix energies can change this flavor
output result.

The initial uncertainty in the momentum of the detector nucleon destroys all
memory of the initial neutrino momentum and of the initial neutrino mass after
the neutrino has been absorbed.  The hand-waving justification of the 
result (\ref{weak3}) uses the
uncertainty principle and says that if we know where the detector is we don't
know its momentum and can't use momentum conservation to determine the mass of
the incident neutrino. The above rigorous justification shows full interference
between the contributions from different neutrino momentum states with the same
energy as long as the product of the momentum difference and the quantum
fluctuations in the initial position of the detector nucleon is negligibly
small in the initial detector state.   

This treatment of the neutrino detector is sufficient to determine the output
of any experiment in which the incident neutrino wave packet is the same well
defined linear combination of mass eigenstates throughout the whole wave
packet. 

\subsection {At what distance is coherence lost?}
  
The above treatment has not comsidered the effects resulting from the 
different
velocities of neutrino 
wave packets with different masses.  
The difference in velocity between components in two wave packets
 $(\delta v)_m$ with the same energy and different mass is just the
difference in velocities $v = p/E$ for states with different momenta and the
same energy,
  \beq{ZZ2b}
(\delta v)_m = {{\partial }\over{\partial p}}\cdot \left(
 {{p}\over{E}}\right)_E \cdot (\delta p)_m =
{{(\delta p)_m }\over{E}}    
     \eeq

The packets will eventually separate and arrive at a remote detector at
different separated time intervals. The detector then sees two separated
probability amplitudes, each giving the probability that the detector observes
a given mass eigenstate. All coherence between the different mass
eigenstates is then lost. The question then arises when and where this occurs;
i.e. at what distance from the source the coherence begin to be lost. Two
different approaches to this problem give the same answer\cite{pnow98}.

1. The centers of the wave packets move apart with the relative velocity
$(\delta v)_m$ given by eq. (\ref{ZZ2b}). Thus the separation $(\delta x)_m$ between
the wave packet centers after a time $t$ when the centers are at a mean distance
$x$ from the source is
\beq{WW5a}
(\delta x)_m=
(\delta v)_m \cdot t= (\delta v)_m \cdot {{x}\over{v}}=
- {{\Delta m^2}\over{2pE}} \cdot {{xE}\over{p}}=
- {{\Delta m^2}\over{2p^2}}\cdot x
\eeq
 
The wave packets will separate when this separation distance is comparable to
the length in space of the wave packet. The uncertainty principle suggests that
the length of the wave packet $(\delta x)_W $ and its spread in momentum space
$(\delta p)_W $ satisfiy the relation
\beq{WW6a}
(\delta x)_W \cdot (\delta p)_W \approx 1/2
\eeq
The ratio of the separation over the length is of order unity when
\beq{WW7a}
\left|{{(\delta x)_m}\over{(\delta x)_W}}\right|  \approx
\left| {{\Delta m^2}\over{p^2}}\right| \cdot (\delta p)_W
 \cdot x \approx 1 \eeq
 
2. Stodolsky\cite{Leo} has suggested that one need not refer to the time
development of the wave packet, but only to the neutrino energy spectrum. The
relative phase $\phi_m(x)$ between the two mass eigenstate waves at a distance
$x$ from the source depends upon the neutrino momentum $p_\nu$ as defined by
the relation (\ref{WW3a}).

Coherence will be lost in the neighborhood of the distance $x$ where the
variation of the phase over the momentum range $(\delta p)_W$ within the wave
packet is of order unity.
For the case of two neutrinos with energy $E$ and mass eigenstates
$m_1$ and $m_2$ the condition that the relative phase variation
$|\delta \phi_m(x)|$ between the two neutrino waves is of order unity
\beq{WW4a}
|\delta \phi_m(x)| = \left|{{\partial \phi_m(x)}\over{\partial p_\nu}}\right|
\delta p_\nu \cdot x = \left| {{\Delta m^2}\over{2p_\nu^2}}\right|
(\delta p)_W \cdot x
\approx 1 \eeq
 
We find that the two approaches give the same condition for loss of coherence.

\section{How incomplete information provides coherence}

\subsection{Bragg Scattering} 

Bragg scattering of photons by a crystal provides an instructive example of
coherence arising from incomplete information on momentum conservation.
Coherence between the photon scattering amplitudes from different atoms in the
crystal produces constructive interference at the Bragg angles and gives peaks
in the angular distribution. When a single photon is scattered from a crystal,
momentum is transferred to the atom in the crystal that scattered the  photon.
If the recoil momentum is detected the atom that scattered the photon is
identified and coherence is destroyed. Coherence arises when quantum mechanics
prevents the measurement of the initial and final momenta of the individual
atoms.

The initial and final states of the crystal are many-particle quantum states
that are eigenstates of the Hamiltonian of the crystal. The dynamics of the
crystal and the interaction with the incident photon allow  elastic scattering,
in which the photon is scattered by a single atom in the crystal but the
quantum state of the crystal is unchanged. This is a purely quantum effect.
Transferring momentum classically to an atom in a crystal must change the
momentum and the motion of the particular atom and allow the identification of
which atom scattered the photon.

The difference produced by quantum mechanics is simply seen in a toy model in
which each atom is bound to its equilibrium position in the crystal by a
harmonic oscillator potential. The atom that scatters the photon is initially
in a definite discrete energy level in the potential. In contrast to the
classical case, the atom cannot absorb the momentum transfer according to the
energy and momentum kinematics of free particles. The final state of the atom
in the potential must be one of the allowed energy levels, and there is a
finite probability that the final state is the same as the initial state. In
this case of elastic scattering, there is no information available on which
atom scattered the photon, and the scattered amplitudes from all scattering
atoms are coherent.

This example shows how amplitudes arising from different processes which would
be classically distinguishable can be coherent. The quantum mechanics of bound
systems can conceal the information which would be classically available from
energy-momentum conservation for free particles.

\subsection{Pion Decay}

This same effect conceals the mass of the neutrino emitted in pion decay. The
initial pion in a beam stop cannot be strictly at rest; it is localized by its
electrostatic interaction with the electric charges in the material where it
was stopped. It is therefore in some kind of energy level of the bound system
and described by a wave function which is a coherent linear combination of
different momentum eigenstates. Measuring the energy of the muon determines the
energy of the emitted neutrino, since the energy of the initial state is
determined. But the momentum of the neutrino is not determined. In a simple toy
model where the initial pion is bound by some external potential, it is
described by a wave function which is a coherent wave packet in momentum space.

When the neutrino strikes a detector, the amplitudes produced by different mass
eigenstates having the same energy and different momenta can be coherent. They
are produced from the different momentum components in the initial pion wave
function which are coherent with a definite relative phase. This can explain
why no electrons are observed at a short distance  from the detector.

If the neutrino amplitudes  produced in this way propagate as free particles,
these considerations determine completely the relative phase between the
amplitudes for neutrinos having the same energy but different masses and
different momenta. The phase change will produce neutrino oscillations with the
same relation between mass differences and phase differences (\ref{WW3a}) that
has been given by the standard treatments.

\section{Time measurements, Momentum and Energy} 

\subsection{The possibility of time measurements}

The preceding analysis does not consider experiments in which the transit time
of the neutrino between source and absorber is measured.
Experiments have been suggested in which the muon emitted together with the
neutrino in a pion decay is observed at the neutrino source and the time  that
the muon is detected is measured precisely along with the time that the muon or
electron is produced by absorbing the neutrino in the detector.  The motivation
is to use some kind of energy-time uncertainty to detect interference between
components  having different energies in the neutrino wave function.

     However, in any realistic detector the quantum fluctuations in the
position of the detector nucleon are small in comparison with the wave length
of the neutrino oscillation. Thus the coherence and the relative phase of the
components in the neutrino wave function having the same energy and different
momenta are preserved. This relative phase completely determines the flavor
output of the detector; i.e. the relative probabilities of producing a muon or
an electron. In all realistic cases where the separation of wave pakets moving
with different velocities is negliible, eqs.(\ref{WW4a}) and (\ref{WW7a}) show
that these probabilities are essentially
independent of energy over the relevant energy range. Thus the relative phases
and coherence between components in the neutrino wave function with different
energies is irrelevant. All energies give the same muon/electron ratio whether
they add coherently or incoherently. Thus time measurements cannot change the
muon/electron ratio observed at the detector.

Thus the flavor output from any time of flight experiment that uses a neutrino
detector that preserves the coherence between states of the same energy and
different momentum is already determined at the single energy level. It is
unaffected by any interference between components of the neutrino wave function
with different energies.

\subsection{The Difference Between Momentum and Energy}

Confusion tends to arise from thinking that momentum and energy should be on
the same footing, particularly since relativity implies that they are
components of the same four vector. But this is only true for isolated free
particles. In any realistic neutrino experiment the neutrino is observed by a
weak interaction with a detector. The detector, in its rest frame before the 
arrival of the neutrino, is in an initial state\cite{Leo} described by a
density matrix in which energy is diagonal and momentum is not. This is the
critical difference between energy and momentum  There is no coherence and no
well-defined relative phase between components in the detector density matrix
with different energies.  But there must be coherence and well defined relative
phases between components with different momenta,  as shown rigorously by eq.
(\ref{weak2}), because we know where the detector is in space and where it
isn't. The form factor (\ref{weak2}) is seen to be negligibly different from
unity as long as the quantum fluctuations in the position of the detector are
small in comparison with the wave length of the oscillation being measured. 

\section{Conclusions}

Coherence between amplitudes produced by neutrinos incident on a detector 
with different masses and the same energy has been shown to follow from the
localization of the detector nucleon within a space interval much smaller than
the wave length of the neutrino oscillation. Decoherence between different mass
eigenstates results from the separation of wave packets moving with different
velocities and is simply described also in terms of the energy dependence of 
the flavor output of a detector. , 

That coherence must exist in neutrinos emitted
from $\pi- \mu$ decay follows from the original Lederman-Schwartz-Steinberger
experiment which saw only muons and no electrons. We now know that at least
two different neutrino mass eigenstates are emitted from pi-mu decay and
that at least one must couple to electrons. The only explanation for the
absence of electrons at the detector is destructive interference from
amplitudes produced by different mass eigenstates.

It is a pleasure to thank Eyal Buks, Maury Goodman, Yuval Grossman, Moty
Heiblum, Yosef Imry, Boris Kayser, Lev Okun, Gilad Perez, David Sprinzak, Ady
Stern, Leo Stodolsky and Lincoln Wolfenstein for helpful discussions and
comments.

%
\catcode`\@=11 
\def\references{ 
\ifpreprintsty \vskip 10ex
%
\hbox to\hsize{\hss \large \refname \hss }\else 
\vskip 24pt \hrule width\hsize \relax \vskip 1.6cm \fi \list 
{\@biblabel {\arabic {enumiv}}}
{\labelwidth \WidestRefLabelThusFar \labelsep 4pt \leftmargin \labelwidth 
\advance \leftmargin \labelsep \ifdim \baselinestretch pt>1 pt 
\parsep 4pt\relax \else \parsep 0pt\relax \fi \itemsep \parsep \usecounter 
{enumiv}\let \p@enumiv \@empty \def \theenumiv {\arabic {enumiv}}}
\let \newblock \relax \sloppy
 \clubpenalty 4000\widowpenalty 4000 \sfcode `\.=1000\relax \ifpreprintsty 
\else \small \fi}
\catcode`\@=12 
{\tighten

} 

\end{document}